# THE SYMMETRY GROUP PARADOX FOR NON-RIGID MOLECULES

B. J. Dalton

Centre for Quantum and Optical Science, Faculty of Science, Engineering and Technology,

Swinburne University of Technology, Melbourne, Victoria 3122, Australia.

**Abstract**

In many situations the energy levels for a quantum system whose Hamiltonian is invariant under a specific symmetry group are split when the Hamiltonian is replaced by a new one with lower symmetry. In non-rigid molecules quantum tunnelling processes allow the molecule to change between different geometrical configurations related by permutations of identical nuclei (or with inversion as well), resulting in the splitting of the energy levels for the rigid molecule case where tunnelling is absent. However, for non-rigid molecules there is apparently a paradoxical situation where although the original rigid molecule energy levels are associated with a symmetry group isomorphic to the point group for the geometrical configuration, the split non-rigid molecule energy levels are associated with a symmetry group consisting of all permutations and inversions related to the quantum tunnelling between configurations, and for which the point group is a sub-group. The resolution of this paradox, where energy level splitting is evidently accompanied by an enlargement of the symmetry group, is the subject of this article.

1. Introduction

In quantum physics there is a standard relationship between the group of symmetry operators that leave the Hamiltonian invariant and the energy eigenvectors for the system, namely that (apart from accidental degeneracy) each set of degenerate energy eigenvectors forms the basis for an irreducible representation of the symmetry group [1]. In cases where the Hamiltonian changes to one whose symmetry group is just a subgroup of the original one, the degenerate energy eigenvectors for the original Hamiltonian are then the basis for a representation of the new and smaller symmetry group, a representation which in general is reducible. Linear combinations of the original eigenvectors can be chosen as basis states for the irreducible representations of the new symmetry group, and these will then be associated with energy eigenvectors for the new, less symmetric Hamiltonian. When the latter only differs from the original Hamiltonian by a small term, its energy levels will be clustered around the original energy levels, so in the *standard situation* a splitting effect associated with a loss of symmetry is involved. Common cases where this occurs are when a free atom is placed in an external electric or magnetic field, where the original rotational symmetry is destroyed.

However, it is well known that for non-rigid molecules the symmetry group Q of feasible permutation-inversions proposed by Longuet-Higgins [2] for classifying the non-rigid molecule energy levels is in general a larger group than the point group R used to classify (see Hougen [3]) the original rigid molecule energy levels (R being a subgroup of Q). The rigid molecule levels split into the non-rigid molecule levels due to the effect of tunnelling processes between configurations of the molecule related by certain permutations (or permutations plus the inversion) of identical nuclei in the molecule that are not included in the point group. However, as several authors have commented (Hougen [4], Petrashen and Trifonov [5], Berry [6]), this situation where the symmetry group becomes larger, is rather different to the standard situation where the splitting of energy levels is associated with symmetry breaking. The review article by Natanson (Ref [7]) also pointed out that this situation is somewhat paradoxical (see P66, par 2).

In this paper we will see how this paradox can be resolved. The approach used to treat non-rigid molecules is based on our earlier paper [8], which showed why the group of feasible permutation –inversions introduced by Longuet-Higgins [2] could be used to classify non-rigid molecule energy levels. There are other approaches such as those involving isodynamic operations [9], isometric groups [10], per-rotations [11]. These, and their relationship to the Longuet-Higgins approach, are discussed in Refs [7], [12], [13]. There are cases where the approaches differ, such as for the Q group and isodynamic approaches in the case of $PF_5$ [13]. The isodynamic operations are sometimes associated with an overall rotation of the molecule, so in this situation the internal and external dynamics are not properly separated.

2. Theory
    2.1 Background to the Paradox

In the standard situation an original Hamiltonian $H_o$ has a symmetry group $G_o$ and a new Hamiltonian H has a symmetry group G. Apart from accidental degeneracies, the degenerate energy eigenvectors of $H_o$ for each energy $E_o$ form the basis for an irreducible representation of $G_o$ whilst the degenerate energy eigenvectors of H for each energy E form the basis for an irreducible representation of G. In the case where G is a subgroup of $G_o$ it follows that the energy eigenvectors of $H_o$ for each energy $E_o$ also form the basis for a representation of G, which in general will be reducible. We can then replace the original degenerate energy eigenvectors of $H_o$ for each energy $E_o$ by linear combinations which now form the basis for irreducible representations of G. In the perturbative case where $H_o$ only differs from H by a small term, the new irreducible basis vectors will be a good approximation to the energy eigenvectors for H. And as they are associated with irreducible representations of its symmetry group G, the irreducible basis vectors will be associated with differing energies E, all clustered around the original energy level $E_o$ for $H_o$. Hence in this case where G is a subgroup of $G_o$ the splitting of the original energy levels is associated with a loss of symmetry in going from $G_o$ to G.

However, the standard situation is not the only one where basis vectors for the irreducible representations of the symmetry group G for a Hamiltonian H can be constructed from the energy eigenvectors of an original Hamiltonian $H_o$ - which only differs from H by a small term. The important feature required is that the degenerate energy eigenvectors for $H_o$ are the basis for a representation of G. Whether they form the basis for an irreducible representation of a symmetry group $G_o$ for $H_o$ is not important. Forming the basis vectors spanning the irreducible representations of G is then associated with the energy levels of H which are clustered around the energy levels $E_o$ for $H_o$. The case where the symmetry group G is a subgroup of $G_o$ is one situation where this feature applies, but it is not the only one. A loss of symmetry is a sufficient condition for energy level splitting, but it is not a necessary condition, as the situation for non-rigid molecules shows.

Before addressing the non-rigid molecule (NRM) paradox we begin with a brief overview of the basic quantum physics of molecular energy eigenstates. The Hamiltonian $H_M$ for a molecule depends on the position and momentum coordinates of all the particles involved, both electrons and nuclei. It may also depend on the spin components for all these particles. The general approach for describing its energy eigenstates is to first separate out the spectroscopically unimportant centre of mass motion, then solve for the energy eigenstates of the fast moving electrons for fixed positions of the slower moving nuclei. For simplicity we will assume that the centre of mass kinetic energy $(P_{cm})^2/2M$ has been removed from the molecular Hamiltonian $H_M$. Within the Born-Oppenheimer approximation (which applies for non-degenerate electronic states) the electronic energy as a function of nuclear positions acts as a potential energy surface, the minima for which define the geometrical structure or equilibrium configuration of the molecule. As we will see, in non-rigid molecules a number of such structures related by permuting identical nuclei between equivalent positions – sometimes inverting the structure as well – will be involved. The $PF_5$ molecule in is lowest electronic state is a case in point. Its geometrical structure is a symmetrical bipramid, the P nucleus being in the middle, two F nuclei occupying the apex positions in the bipyramid, and the other three F nuclei forming an equilateral triangle in the equatorial plane. Clearly, the electronic energy would still be a minimum whichever two of the five identical F nuclei occupied the apex positions and whichever three occupied the equatorial positions. Furthermore, all the F nuclei could be inverted through the position of the P nucleus, and the potential energy would still be a minimum. As the nuclei also have momentum coordinates, they undergo a vibrational motion around each equilibrium geometrical structure, which also undergoes a slower rotational motion. There are of course a number of distinct vibrational modes involved and in general the vibrational and rotational motions are coupled. Overall, if nuclear spin interactions and the rotation-vibration couplings are ignored, the overall energy is the sum of the electronic energy for the equilibrium structure, the vibrational energy and the rotational energy. Both the vibrational and rotational energy contributions are quantised, the former being determined by the number of quanta in each of the vibrational modes and their frequencies, the latter depending on the overall orbital angular momentum quantum numbers and the moments of inertia associated with the geometrical structure considered as a rigid nuclear frame. Cases involving degenerate electronic states lead to Jahn-Teller effects coupling the electronic and vibrational dynamics, but we will not consider this situation. Finally, we point out that the NRM effects are associated with changes to the so-called rovibronic (rotational and vibrational) contributions to the energy levels due to the effects of quantum tunnelling processes between different configurations – in the simplest case, related by permutations of the identical nuclei (plus the inversion as well in some cases). However, as Longuet-Higgins [2] pointed out, to consider these effects we need to take note of the fact that (apart from overall rotational and translational symmetry and ignoring the electrons) the actual symmetry group for the molecular Hamiltonian $H_M$ is the full permutation-inversion group P containing all the permutations of the position, momentum and spin components of the identical nuclei in the molecule, or such permutations combined with the inversion.

### 2.2 The Hamiltonians $H_R$ and $H_{NRM}$

In considering the paradox (see footnote [14]), the original Hamiltonian $H_o$ will be replaced by $H_R$ which describes the rigid (though rotating and vibrating) molecule. For a specific original configuration, the rovibronic states |(robivib) $E_o$ LM i> ($|\Gamma$ i> for short), where I = 1, 2, …, $d_E$ lists states for the rigid molecule with same energy $E_o$, L,M are the angular momentum and magnetic quantum numbers) are eigenstates of a Born-Oppenheimer Hamiltonian $h_R$ (to be related to $H_R$) describing small vibrations around the rotating original configuration. These states are associated with irreducible representations of the molecular point group R, with R(g) $|\Gamma$ i> = $\sum_{(j)} S_{j;I} |\Gamma$ j>. The point group (order p) is isomorphic to a subgroup of the full permutation-inversion group P, as Hougen [3] showed in the early 1960's. The degeneracy $d_E$ would therefore correspond to the dimension of one of the irreducible representations of R. As all its elements correspond to permutation-inversions - which leave the molecular Hamiltonian $H_M$ invariant, R is clearly eligible to be the symmetry group for $h_R$.

Also, to be similar to the standard situation the Hamiltonian H will be replaced by a Hamiltonian $H_{NRM}$ describing the non-rigid molecule - where feasible changes of configuration from the original configuration are involved, and where the symmetry group for $H_{NRM}$ is Q (the Longuet-Higgins group [2] – order f), which is also a subgroup of P - the point group R being isomorphic to subgroup of Q. It has been known since the mid 1960's (Dalton [8]) that if the Q group is divided up into the cosets (Dalton [8], Watson [15]) of its subgroup R, then (as will be explained below) the coset generators $G_1$, $G_2$, …, $G_r$, … , $G_{f/p}$ define configurations linked to the original configuration (defined by $G_1$ - the unit permutation-inversion) via feasible NRM tunnelling processes. The set of states $G_r|\Gamma$ i> for I = 1, 2, …, $d_E$ and for each r = 1, 2,…,f/p specify the energy eigenstates for the Hamiltonian $G_r\, h_R\, G_r^{-1}$ and describe the rigid molecule rovibronic states for the r th configuration. For each configuration r the $G_r|\Gamma$ i> for I = 1, 2, …, $d_E$ would be associated with irreducible representations for the groups $G_r\, R(g)\, G_r^{-1}$ – each of which is of course isomorphic to the point group. As each set of $G_r|\Gamma$ i> for I = 1, 2, …, $d_E$ specifies the same rigid molecule energy levels, then the overall set for all of the configurations are the energy eigenvectors for an overall rigid molecule Hamiltonian $H_R$ - which will be analogous to $H_o$ in the standard situation. We can construct $H_R$ from its matrix for the Born-Oppenheimer Hamiltonians $G_r \, h_R \, G_r^{-1}$ with respect to the basis of $G_r|\Gamma\, i\rangle$ for I = 1, 2, ..., $d_E$ and r = 1, 2,...,f/p. This matrix would consist of identical diagonal blocks for the different configurations with the off-diagonal elements for r not equal to s all being zero. Each diagonal block is also diagonal with rigid molecule energies $E_o$ along the diagonal.

Analogous to the standard situation, we choose the energy eigenvectors $G_r|\Gamma\, i\rangle$ for I = 1, 2, ..., $d_E$ and r = 1, 2,...,f/p for $H_R$ (analogous to $H_o$) to construct the energy eigenvectors for $H_{NRM}$ (analogous to H). The matrix for $H_{NRM}$ is defined by taking the matrix elements of the molecular Hamiltonian $H_M$ between these states. It also contains identical diagonal blocks for the different states $G_r|\Gamma\, i\rangle$ i = 1, 2, .., $d_E$ for any particular r, but due to NRM processes linking the different configurations the off-diagonal elements for r not equal to s are now no longer zero – as was the case for $H_R$. We can of course express the Hamiltonians $H_R$ and for $H_{NRM}$ in operator form, but this is not necessary. Note that $H_R$ and $H_{NRM}$ are different, just as $H_o$ and H were for the standard situation - so there is no reason to expect them to have the same symmetry group. However, because of the way the symmetry group Q was constructed, we can show that the energy eigenvectors $G_r|\Gamma\, i\rangle$ for $H_R$ form a basis for a *reducible* representation of Q – just as the energy eigenvectors for $H_o$ were a basis for a reducible representation of G.

It should be pointed out that neither $H_R$ nor $H_{NRM}$ is the same as the molecular Hamiltonian $H_M$. The first reason is that both these Hamiltonians have been constructed from matrix elements associated with only a limited set of basis vectors, namely the $G_r|\Gamma\, i\rangle$ for I = 1, 2, ..., $d_E$ and r = 1 ,2,...,f/p. Apart from the special case (such as in $PF_5$) where the Q group is the same as the full permutation-inversion group P (order n), there are further configurations not listed by the coset generators $G_1$, $G_2$, ..., $G_r$, ..., $G_{f/p}$, which are *not* linked to $G_1$ via feasible NRM processes. As will be explained later, these are associated with generators of the cosets of the point group listed as $F_t \, G_r$ (with t = 1, 2, .., n/f and r = 1, 2, ..., f/p), where $F_1$, $F_2$, ... , $F_{(n/f)}$ are the generators of the cosets of the Q group within P. These are associated with their own sets of rigid molecule states $F_t \, G_r|\Gamma\, i\rangle$ for I = 1, 2, ..., $d_E$. Since the rigid molecule states for $F_2$, ... , $F_{(n/f)}$ have *not* been included when $H_R$ and $H_{NRM}$ were constructed, neither could be the same as $H_M$. The second reason is that whilst $H_{NRM}$ does involve matrix elements of $H_M$, the same is not the case for $H_R$, where matrix elements for Born-Oppenheimer Hamiltonians $G_r \, h_R \, G_r^{-1}$ were involved.

### 2.3 The Longuet-Higgins Q Group and its Irreducible Representations

To justify the statements in the previous section about the role of Q, we need to consider first how the Q group is determined. The construction of the Q group is based on noting that the non-negligible Hamiltonian NRM matrix elements link the original configuration (specified by coset generator $G_1$ for the point group) either directly or indirectly to a set of configurations specified by coset generators $G_2$, $G_3$, ..., $G_{f/p}$. Thus if $G_2$, $G_3$, ..., $G_r$, ...., $G_t$, ..., $G_d$ specify the configurations *directly* linked to $G_1$, we find a set of related matrix elements for the molecular Hamiltonian

$\langle\Gamma\, i|(G_1)^{-1} H_M (G_r R(g)) |\Delta\, j\rangle = \langle\Gamma\, i| (G_s R(h))^{-1} H_M (G_s R(h)) (G_r R(g))|\Delta\, j\rangle$

$= \langle\Gamma\, i|((G_t R(i)) (G_s R(h)))^{-1} H_M (G_t R(i)) (G_s R(h)) (G_r R(g))|\Delta\, j\rangle =...$ (1)

where since permutation-inversion operators are unitary $(G_r R(g))^{-1}$ etc is equal to the Hermitian adjoint of $(G_r R(g))$ etc. The r, s, t, etc and the g, h, i, .., can be in any order. This result follows from $H_M$ being invariant under any permutation-inversion.

It is easy to see that the set of *all* $(G_r R(g))$, $(G_s R(h)) (G_r R(g))$, $(G_t R(i)) (G_s R(h)) (G_r R(g))$ etc forms a group Q  (see [8]), of which the point group is a subgroup - and which in turn is generally a subgroup of the full permutation-inversion group P. Also, each one of these products must be in one of the cosets $G_1$, $G_2$, $G_3$, …, $G_{f/p}$ of R within Q, and from Eq (1) it is clear that the rigid molecule states $G_r|\Gamma$ i> specified by these generators are either directly or indirectly linked by non-zero NRM matrix elements. Also (as we will see), the group Q is the symmetry group for $H_{NRM}$ since $H_{NRM}$ has a diagonal matrix in terms of the irreducible representations of Q. Since all its elements are permutation-inversions - which leave the molecular Hamiltonian $H_M$ invariant, Q is clearly eligible to be the symmetry group for $H_{NRM}$.

To show that the energy eigenvectors for $H_R$ form the basis for a representation of Q we see that as Q(h) $G_s$ must be a member of Q it can be written in the form $G_u R(g)$. Hence

Q(h) $G_s |\Gamma$ i>= $G_u R(g)|\Gamma$ i>, which is just a linear combination of the $G_r|\Gamma$ j>. We can write this as Q(h) $G_s |\Gamma$ i>= $\sum_{(r,j)}$ $S_{rj;si}$ (h) $G_r |\Gamma$ j>, where $S_{rj;si}$(h)=$\delta_{r;u} S_{j;l}$(g). This enables us to determine the character function $\chi$(h)= $\sum_{(s,i)}$ $S_{si;si}$(h) for this representation of Q, and to then calculate the number of times $M_\lambda$ that a particular irreducible representation $\lambda$ occurs. This is given by

$M_\lambda$ = (1/f) ) $\sum_{(h)}$ $\chi^\lambda$(h)$^*$ $\sum_{(s,i)}$ $S_{si;si}$(h) = (1/f) ) $\sum_{(s,i)}$ $\sum_{(h)}$ $\chi^\lambda$(h)$^*$ $S_{si;si}$(h). For non-zero $S_{si;si}$(h) we require Q(h) $G_s$ = $G_s$ R(g) in which case $S_{si;si}$(h) = $S_{i;i}$(g) and Q(h) = $G_s$ R(g) $G_s^{-1}$. Since $\chi^\lambda$( $G_s$ R(g) $G_s^{-1}$ ) = $\chi^\lambda$( R(g) ) independent of s, we can sum over s to give f/p and replace the sum over h by a sum over the point group elements g. We thus find that

$M_\lambda$ = (1/p) ) $\sum_{(g)}$ $\chi^\lambda$(g)$^*$ $\sum_{(i)}$ $S_{i;i}$(g)                                                                                  (2)

which just involves a sum over point group elements. This expression (which was first stated by Watson [15] and Dalton [8] in the mid 1960s) determines how the rigid molecule levels are split into the NRM energy levels, for which the energy eigenvectors form the basis for irreducible representations of the NRM symmetry group Q – just as in the standard situation the energy eigenvectors for H formed the basis for irreducible representations of G.

The irreducible basis vectors for the symmetry group Q formed as linear combinations of the entire set of $G_r|\Gamma$ i> for i=1, 2, …, $d_E$  and r-1,2,…,f/p are then of the form  (see [8])

$|\Gamma \lambda\mu j>=\sum_{(r,i)}$ C(ri; $\lambda\mu$: $\Gamma$) $G_r|\Gamma$ i>,                                                                                                    (3)

where $\mu$ = 1, 2, …, $M_\lambda$, enumerates different irreducible representation of species $\lambda$, j = 1, 2, …, $d_\lambda$ denotes vectors of the species $\lambda$ that transform among themselves under the operations in Q, and $d_\lambda$ is the dimension for irreducible representation $\lambda$. The transformation matrix is denoted $S^\lambda_{j;l}$(h). The matrix of the molecular Hamiltonian $H_M$ with respect to this basis is diagonal. Since Q(h)$^{-1}$ $H_M$ Q(h) =$H_M$ for any element Q(h) in Q we see that

<$\Gamma \lambda\mu i|H_M |\Delta \nu\eta l$> = <$\Gamma \lambda\mu i|$ Q(h)$^{-1}$ $H_M$ Q(h) $|\Delta \nu\eta l$> = $\sum_{(k,j)}$ $S^\lambda_{j;i}$(h)$^*$ $S^\nu_{k;l}$(h) <$\Gamma \lambda\mu j|H_M |\Delta \nu\eta k$>.

If we then average over Q group elements Q(h) and use the well-known orthogonality theorem (see Ref. [1]) for the irreducible transformation matrices we find that

$\langle \Gamma \lambda\mu i | H_M | \Delta \nu\eta l \rangle = \delta_{\lambda\nu} \delta_{i;l} (1/d_\lambda) \Sigma_{(j)} \langle \Gamma \lambda\mu j | H_M | \Delta \lambda\eta j \rangle$

$\langle \Gamma \lambda\mu i | H_M | \Gamma \nu\eta l \rangle = \delta_{\lambda\nu} \delta_{i;l} \delta_{\mu\eta} (1/d_\lambda) \Sigma_{(j)} \langle \Gamma \lambda\mu j | H_M | \Gamma \lambda\mu j \rangle$ (4)

The last equation follows because for $\Gamma = \Delta$ we could also have chosen the different irreducible representations $\mu$, $\eta$ for species $\lambda$ to give diagonal Hamiltonian matrix elements in $\mu$, $\eta$. The matrix elements $\langle \Gamma \lambda\mu j | H_M | \Delta \lambda\eta j \rangle$ depend ultimately on the non-zero $\langle \Gamma i | G_1^{-1} H_M G_r | \Delta k \rangle$ that define the NRM tunnelling processes. For the situation where $\Gamma = \Delta$ and the non-rigid molecule matrix elements are small compared to the rigid molecule energy level spacings, the splittings of the original rigid molecule levels associated with species $\lambda$ for the Q group are obtained from the NRM energy levels $E_{\lambda\mu}(\Gamma)$ – which are clustered around $E_o$

$E_{\lambda\mu}(\Gamma) = (1/d_\lambda) \Sigma_{(j)} \langle \Gamma \lambda\mu j | H_M | \Gamma \lambda\mu j \rangle$ (5)

### 2.4 Resolving the Paradox

So far then, the situation for non-rigid molecules mirrors that for the standard situation. One way to resolve the paradox would therefore be to show that the standard situation fully applies, and that the energy eigenvectors for $H_R$ for any particular energy $E_o$ *are* the basis for an irreducible representation of a suitably defined symmetry group $\mathbb{R}$ for $H_R$. In addition, assuming such a symmetry group can be found, then for the standard situation to apply the symmetry group Q for $H_{NRM}$ should be a *subgroup* of $\mathbb{R}$. So we would need to identify a symmetry group $\mathbb{R}$ for $H_R$ (which is analogous to the symmetry group $G_o$ for $H_o$) such that the entire set of $G_r | \Gamma i \rangle$ for $I = 1, 2, ..., d_E$ and $r = 1, 2, ..., f/p$ containing $d_E \times (f/p)$ vectors is the basis for such an irreducible representation. At present, there is no clear proof that such group $\mathbb{R}$ can be identified. As has been noted previously, both the point group and the Q group were possible symmetry groups because all of their elements were members of the full permutation-inversion group P, which is the actual symmetry group for the molecular Hamiltonian $H_M$. The symmetry group $\mathbb{R}$ for $H_R$ should also be made up of elements in P, but (as will now be seen) this is not possible in general. As the sum of the squares of the dimension of each irreducible representation equals the order of the group, it follows that if $\mathbb{R}$ exists its order must be at least $(d_E \times (f/p))^2$. We can then at least be sure that it is not isomorphic with the point group - whose order is p, so if $\mathbb{R}$ exists we would have resolved the problem of it being incorrectly identified as being the point group – as in the paradox. However, for the paradox to be resolved $\mathbb{R}$ should be larger than both the point group and the Q group used to classify the NRM levels, but at the same time be made up of elements in P. An example shows this is not possible in general. In the case of Berry pseudo-rotation [16] in $PF_5$, the $D_{3h}$ point group has p = 12 elements and the Q group is the same as the full permutation-inversion group P with n = f = 2x(5!) = 240 elements (see Ref. [17]). For rovibronic states of $D_{3h}$ point group symmetry E, we have $d_E = 2$, so the required symmetry group $\mathbb{R}$ would need to have at least $(2x(240/12))^2 = 1600$ elements in it if the set of $G_r | \Gamma i \rangle$ for $I = 1, 2, ..., d_E$ and $r = 1, 2, ..., f/p$ eigenstates are to form the basis for an *irreducible* representation of $\mathbb{R}$. Since the entire permutation-inversion group P only contains 240 elements and any genuine symmetry group should be a sub-group of P, the particular example of PF$_5$ shows that in general no such symmetry group ℝ exists for which Q is a subgroup in order to conform to the standard situation - where splitting is a result of symmetry reduction.

So what is the answer to the paradox ? Here are two answers.

The first is to say that the paradox is not a real paradox at all, and only exists in the minds of those who have a preconceived notion that energy level splitting *only* occurs when the original (and approximate) energy eigenvectors are associated with an irreducible representation of a symmetry group ℝ for which the symmetry group Q for the final (and exact) energy eigenvectors is a subgroup. As we have seen, all that is necessary is that the approximate rigid molecule energy eigenvectors associated with configurations linked by feasible tunnelling processes are the basis for a (reducible) representation of Q, and an ℝ for which the original energy eigenvectors form an *irreducible* representation) does not even need to exist. Forming the irreducible basis vectors for Q from the original energy eigenvectors then leads to the only important required outcome, namely that the exact energy eigenvectors form the basis for an irreducible representation of Q.

The second (which is perhaps the more satisfactory of the two) is to recognise that for any choice of energy resolution for identifying the rigid or non-rigid molecule energy levels, the true energy eigenvectors allowing for nuclear spin must provide the basis for irreducible representations of the full permutation-inversion group P, since this is the actual symmetry group for the molecular Hamiltonian H$_M$. However, since Q is a subgroup of P we can identify generators F$_1$, F$_2$, … , F$_{(n/f)}$ of the cosets of the Q group within P, so that with G$_1$ , G$_2$ , G$_3$ , …, G$_{f/p}$ as the generators of the cosets of R within Q, the set of generators for cosets of R within P can be written as F$_t$ G$_r$ (with t = 1, 2, .., n/f and r = 1, 2, …, f/p). For F$_t$ not equal to the unit permutation-inversion, the basis vectors F$_t$ G$_r$|Γ i> for i = 1, 2, …, d$_E$ would then be approximate energy eigenvectors for rovibronic states associated with the remaining configurations which are *not* linked to the original configuration (where F$_1$ G$_1$ equals the unit permutation-inversion) by feasible NRM tunnelling processes. These approximate energy eigenvectors must also be taken into account when the true energy eigenvectors associated with H$_M$ are formulated. Furthermore, according to the symmetrisation principle, the final irreducible representation for the full permutation-inversion group is restricted to be one where applying a permutation operator to the true energy eigenvector must result in either +1 or -1, according as the permutation operator for identical fermion nuclei is either even or odd (and +1 for any permutation of identical boson nuclei). Also, when the inversion operator is applied the result can be either +1 or -1 corresponding to whether the energy eigenvector is even or odd under inversion. There are just two such irreducible representations for P, referred to as S+ and S-. Both are symmetrized as just described when applying a permutation operator, the + or – distinguishing them by their evenness or oddness under inversion. To obtain the true energy eigenvectors for the non-rigid molecule the irreducible basis vectors |Γ λμj> described previously, and obtained by considering only the rovibronic motion in the non-rigid molecule for configurations associated with Q, will first have to be combined with suitable nuclear spin states, and then the presence of NRM states $F_t |\Gamma \lambda\mu j\rangle$ involving configurations whose coset generators (t = 2, 3, ..., n/f) are not in the symmetry group Q must be allowed for.

The procedure is set out in Ref. [8] and only the results presented here. The details differ depending on whether Q consists of half permutations and half permutation-inversions (Case (b)) or whether it consists only of permutations (Case (a)).

For Case (a) we first form irreducible basis vectors $|(spin) \lambda(S) \theta j\rangle$ in spin space, where the transformation matrix for the $\lambda(S)$ irreducible representation for Q is given by $\varepsilon(h) S^\lambda_{j;i}(h)^*$, where $\varepsilon(h) = +1, -1$ according as the element Q(h) involves an even or odd permutation, and $\theta = 1, 2, ..,$ $m_{\lambda(S)}$ and $j = 1, 2, ..., d_\lambda$ have the usual meanings. The vectors that transform according to the S+ and S- irreducible representations of the full permutation-inversion group P are given by

$|(rovib-spin) S+; \Gamma \lambda\mu\theta\rangle = \Sigma_{(t)} \varepsilon(t) F_t \{ \Sigma_{(j)} |\Gamma \lambda\mu j\rangle |(spin) \lambda(S) \theta j\rangle \}$

$|(rovib-spin) S-; \Gamma \lambda\mu\theta\rangle = \Sigma_{(t)} u(t) \varepsilon(t) F_t \{ \Sigma_{(j)} |\Gamma \lambda\mu j\rangle |(spin) \lambda(S) \theta j\rangle \}$ (6)

where $\varepsilon(t) = +1, -1$ according as the element $F_t$ involves an even or odd permutation, and $u(t) = -1, +1$ according as the element $F_t$ involves, does not involve the inversion. Both the even and the odd states have the same NRM energy. The number of times $m_{\lambda(S)}$ that the $\lambda(S)$ irreducible representation occurs when the nuclear spin space is reduced determines the overall statistical weight, which is 2 $m_{\lambda(S)}$.

For Case (b) we first form irreducible basis vectors $|(spin) \lambda(S+) \theta j\rangle$ in spin space, where the transformation matrix for the $\lambda(S+)$ irreducible representation for Q is given by $\varepsilon(h) S^\lambda_{j;l}(h)^*$ and $\theta = 1, 2, .., m_{\lambda(S+)}$, and we also form irreducible basis vectors $|(spin) \lambda(S-) \theta j\rangle$ in spin space, where the transformation matrix for the $\lambda(S-)$ irreducible representation for Q is given by $u(h) \varepsilon(h) S^\lambda_{j;l}(h)^*$ and $\theta = 1, 2, .., m_{\lambda(S-)}$. The other symbols have the usual meanings. The vectors that transform according to the S+ and S- irreducible representations of the full permutation-inversion group P are given by

$|(rovib-spin) S+; \Gamma \lambda\mu\theta\rangle = \Sigma_{(t)} \varepsilon(t) F_t \{ \Sigma_{(j)} |\Gamma \lambda\mu j\rangle |(spin) \lambda(S+) \theta j\rangle \}$

$|(rovib-spin) S-; \Gamma \lambda\mu\theta\rangle = \Sigma_{(t)} u(t) \varepsilon(t) F_t \{ \Sigma_{(j)} |\Gamma \lambda\mu j\rangle |(spin) \lambda(S-) \theta j\rangle \}$ (7)

where $\varepsilon(t) = +1, -1$ and $u(t) = -1, +1$ have their previous meanings. Both the even and the odd states have the same NRM energy. The number of times $m_{\lambda(S+,-)}$ that the $\lambda(S+,-)$ irreducible representations occur when the nuclear spin space is reduced determines the statistical weight, which is ($m_{\lambda(S+)} + m_{\lambda(S-)}$).

The results in Eqs. (6) and (7) for the true energy eigenvectors present us with perhaps the most satisfactory resolution to the paradox. This is that the actual symmetry group for non-rigid molecules is *always* the full permutation-inversion group P, and that the true energy eigenvectors transform according to either the S+ or S- irreducible representations of P. The *same* conclusion would also apply even if NRM tunnelling processes were entirely absent, as in rigid molecules. In this case Q would just be replaced by R and the generators $F_t$ would just be those for *all* the cosets of R within P. These would be equivalent to all the $F_t G_r$ for t = 1, 2, 3, ..., n/f and r = 1, 2, ..., f/p – and which could be re-listed as $G_\alpha$ with $\alpha = 1, 2, ..., n/p$. This point cannot be overemphasised – even for the rigid molecule case the true energy eigenvectors transform as one dimensional irreducible representations S+ and S – of the full permutation-inversion group. There is no change to the true symmetry group in going from the rigid molecule to the NRM situation.

This is not to discount the important roles both the point group R and the Longuet-Higgins Q group play as being the *effective* symmetry groups involved for determining the splitting of the rigid molecule energy levels, the statistical weights of the NRM energy levels and the selection rules for radiative transitions (see Ref. [8] for details) between them.

However, that the symmetry group is always P is really not that surprising. The molecular symmetry group P for the Hamiltonian $H_M$ does not change in any way if the molecule changes from one where NRM effects are unobservable to one where the NRM effects result in an observable splitting of the original rigid molecule levels. Such a transition could be accomplished by merely changing parameters in the molecular Hamiltonian, such as the nuclear masses - since quantum tunnelling rates vary dramatically with the masses of the particles involved. It is therefore not surprising that the fundamental symmetry group for the true energy levels remains P, irrespective of whether the molecule is rigid or non-rigid.

Another way of showing that P is always the true symmetry group is based on noting that if the molecular energy levels were resolved using higher and higher resolution, then eventually the splitting effects of *all* possible NRM tunnelling effects would be seen. In this case the Longuet-Higgins Q group is the full permutation-inversion group P, all generators as $G_\alpha$ with $\alpha = 1, 2, ..., n/p$ would be involved in determining the irreducible basis vectors $|\Gamma \lambda\mu j>$ for the spatial factor in the energy eigenvectors (with $\lambda$ now an irreducible representation of P). In forming the true energy eigenvectors as in Eq (7) after including the nuclear spin states (note that since Q = P, the Case (b) situation applies), there would only be one term with $F_t$ given by the unit permutation-inversion for the two irreducible representations $|$(rovib-spin) S+; $\Gamma \lambda\mu\theta>$ and $|$(rovib-spin) S-; $\Gamma \lambda\mu\theta>$ . So in this limit it is clear that the irreducible representations for the full permutation-inversion group P are involved, not only for the final space-spin energy eigenvectors, but also in determining the spatial factors $|\Gamma \lambda\mu j>$.

Two answers to the paradox have now been presented. A third answer was given by Natanson (Ref. [7]) and essentially involved proposing hidden symmetry groups $D^{glob}$ for the rigid molecule and NRM situations. First, a semidirect product $T^{glob}$ of the direct product of groups associated with mapping states for the various configurations onto themselves, together with the group of all permutations was introduced. The hidden symmetry group $D^{glob}$ proposed then involved the semidirect product of a group $\Delta$ based on certain additional symmetry operations applied in the various configurations that multiply functions by -1, with the previous group $T^{glob}$. For the rigid molecule states the hidden symmetry group is $D_{rig}$ whilst that for NRM states it is $D_{nonrig}$, with the latter being a subgroup of the former (see Eq 2.5 in Ref. [7]) - as in the standard situation. Also, the full permutation-inversion group is a subgroup of $D_{nonrig}$. However, no proof was given to show that the degenerate states for the rigid molecule situation are the basis for an irreducible representation of $D_{rig}$, as would be needed to resolve the paradox.

### 2.5 Other Situations

The previous sections dealt with the simplest case, namely where only configurations with the same geometrical shape were involved. However, there are NRM cases involving tunnelling between configurations which have different shapes, such as for the butyl ion $C_4H_9^{+\cdot}$, where there are four different isomers involved. Nevertheless, the Longuet-Higgins Q group approach still applies, basically because a NRM process between an original configuration and one of a different shape acts as a pathway for a NRM process to a further configuration with the same geometrical shape as the original, but related to it by a permutation-inversion. This situation together with cases involving linear geometrical shapes (such as in $H_2F_2$) are discussed in Ref. [18], where again the Q group applies. Hyperfine splitting effects on the NRM energy levels due to weak nuclear spin interactions are treated in Ref. [19]. In this case the hyperfine levels are classified in accordance with the F=J+I quantum number, along with parity.

### 3. Conclusion

If a detailed examination is made of how the true energy eigenvectors for non-rigid molecules is carried out, including how both the point group for the molecular configuration and the Longuet-Higgins symmetry group Q of feasible permutation-inversions appear in the theory, it is possible to resolve the apparent paradox of the splitting of the original rigid molecule levels being associated with an increase rather than a decrease of symmetry.

In the present article the paradox has been resolved in two ways. The first is that **the paradox is not a real paradox at all, and only exists in the minds of those who have preconceived notion that energy level splitting *only* occurs when the original (and approximate) energy eigenvectors are associated with an irreducible representation of a symmetry group $\mathbb{R}$ for which the symmetry group Q for the final (and exact) energy eigenvectors is a subgroup. As we have seen, it is impossible to construct such a symmetry group $\mathbb{R}$ in the NRM case that consists of permutation-inversion operators that leave the molecular Hamiltonian invariant. Also, in determining the true energy eigenvectors this group is also un-necessary, since the entire set of approximate energy eigenvectors for rigid molecule configurations linked by feasible non-rigid molecule permutation-inversion processes satisfies the key requirement of forming a basis for a representation of Q. This is sufficient to show why the split levels are associated with irreducible representations of Q. The second way of resolving the paradox is to focus on the true energy eigenvectors taking into account the rigid molecule states for all possible configurations (irrespective of whether they are linked by feasible permutation-inversions or not) together with the nuclear spin states. If this is done, it then follows that the true energy eigenvectors are one of two irreducible one dimensional representations (S+ and S-) of the full permutation-inversion group P. Both are symmetrised with respect to permutations of the identical nuclei, one being even the other odd with respect to inversion. The same applies also in the rigid molecule situation, so there is no change in the true symmetry group between the rigid molecule and NRM cases. There is no symmetry breaking when the rigid molecule levels split. Thus, the symmetry group for the true energy eigenvectors is always the permutation-inversion group P – which is not unexpected since the molecular Hamiltonian $H_M$ has P as its symmetry group. This is of course not to minimise the role that the Q group plays in determining the splitting of the original rigid molecule levels, together with their statistical weights and the selection rules for radiative transitions. Nevertheless, irrespective of whether non-rigid molecule processes are occurring (or not), even though Q (or R) acts as the *effective* symmetry group for these purposes, it is really not the true symmetry group.

**Acknowledgement** The author acknowledges correspondence with G. A. Natanson, whose 1985 paper drew his attention to the symmetry group paradox.